# High-precision molecular dynamics simulation of $UO_2$–$PuO_2$: pair potentials comparison in $UO_2$

S.I. Potashnikov[a], A.S. Boyarchenkov[a], K.A. Nekrasov[a], A.Ya. Kupryazhkin[a]

[a] Ural Federal University, 620002, Mira street 19, Yekaterinburg, Russia

potashnikov@gmail.com  boyarchenkov@gmail.com  kirillnkr@mail.ru  kupr@dpt.ustu.ru

**Abstract**

Our series of articles is devoted to high-precision molecular dynamics simulation of mixed actinide-oxide (MOX) fuel in the approximation of rigid ions and pair interactions (RIPI) using high-performance graphics processors (GPU). In this first article 10 most recent and widely used interatomic sets of pair potentials (SPP) are assessed by reproduction of solid phase properties of uranium dioxide ($UO_2$) – temperature dependences of the lattice constant, bulk modulus, enthalpy and heat capacity. Measurements were performed with 1K accuracy in a wide temperature range from 300K up to melting points. The best results are demonstrated by two recent SPPs MOX-07 and Yakub-09, which both had been fitted to the recommended thermal expansion in the range of temperatures 300–3100K. They reproduce the experimental data better than the widely used SPPs Basak-03 and Morelon-03 at temperatures above 2500K.

## *1. Introduction*

The intensive development of nuclear power technology places high demand for the nuclear reactor materials. One of the most critical parts is the fuel element, which is made on the basis of actinide-oxide compounds: $UO_2$ (~95% of all fuel), $PuO_2$, $ThO_2$, etc. [1]. Of great interest is mixed oxide (MOX) fuel, with regard to the closed fuel cycle, nuclear proliferation and effective utilization of weapon-grade plutonium (Pu-239). MOX-fuel of $(U,Pu)O_2$ type have been already used in some commercial light water reactors (LWR) and experimental fast breeder reactors (FBR).

Forecasting the behavior of MOX-fuel in fabrication, operation and recycling processes is a prerequisite for safe and effective nuclear energy. Moreover, its characteristics are of paramount importance, considering the danger of "Loss Of Coolant Accident" (LOCA), when fuel melt down occurs [2].

The necessary experiments with high temperatures (~3000K), pressures (~1–10 GPa) and radiation levels (especially for toxic plutonium) are not always possible, because all instruments have their limits of durability. In such extreme conditions, molecular dynamics (MD) simulation is often the best way to obtain the necessary information.

Because of its computational complexity MD-simulation of such many-electron elements as actinides is usually performed in the approximations of rigid ions and pair interactions (RIPI) instead of first principles. In this case all structural and transport properties of the model are completely determined by the chosen set of pair potentials (SPP).

Parameterization of SPP can be based on *ab initio* (from first principles) calculations or the known experimental data (empirical potentials). At present, *ab initio* calculations of actinide-oxide fuel are performed in static approximations of density functional theory (see review in [3]) without particle dynamics and so without taking into account the kinetic (thermal) properties of the model. Therefore, they do not allow analysis of the temperature dependences of characteristics. In addition, the work [4] notes that even with the same approximations and close numerical parameters a discrepancy between the results of *ab initio* calculations has been quite large so far.

At the same time, adequacy of empirical parameterization of SPP is improving with the development of computational tools and refinement of experimental data. The parameterization has evolved from the simplest calculations of binding energy, dielectric, elastic properties [5] [6] and phonon spectra [7] in the harmonic oscillators approximation to the lattice statics calculations of the point defects formation energies [8] and, finally, to self-consistent MD-simulation of temperature dependences, which takes into account the kinetic (thermal) effects [9] [10] [11] [12] [13].

Nevertheless, in practice a choice of the most suitable SPP lacks of comparative surveys, because authors of new potentials tend to compare them only with experimental data and one or two predecessors. The first (and, to our knowledge, the only) broad comparative review is the work of Govers et al. in two parts: with static [14] and MD calculations [15], covering more than 20 SPPs for uranium dioxide ($UO_2$), proposed over the past 40 years. Unfortunately, in the MD part authors did not simulate diffusion and the temperature dependences were measured with a too coarse step of 100–200K and only up to 3000K (which, in particular, cannot be used as a source of premelting and melting characteristics comparison). Besides, the parameters of Yamada's, Basak's and Arima's potentials were incorrectly translated by authors from kJ/mol to eV (per molecule) – with a coefficient of 96.441 instead of the standard 96.485 [16], which also decreased the accuracy (because

SPPs are sensitive even to small changes of parameters). In particular, MD simulations in the review [15] were performed on a system of 768 ions, for which using for example Basak-03 potentials with parameters from [14] instead of the original causes displacement of λ-peaks by ~100K and deviation up to $0.4*10^6$/K and up to 0.026 kJ/(mol*K) for linear thermal expansion coefficient (LTEC) and heat capacity temperature dependences.

Regarding MD-simulation of mixed $(U,Pu)O_2$ fuel, we are aware of only two works [9] [11] (during the preparation of this article another SPP of Tiwary et al. was published [17], which are going to be considered in our future survey of $PuO_2$ and MOX potentials), in which authors proposed SPPs in RIPI approximation, and there are no any comparative reviews on this compound.

Therefore, we have set ourselves the task of accurate comparison of thermophysical properties of the most recent and widely used SPPs, giving special attention to phase transitions (Bredig superionic transition and melting), mass transport mechanisms (including self-diffusion of slow-moving cations) and nanoscopic crystals with surface. Due to the large amount of interesting results we have divided them into several publications, and this article focuses on simulation of the $UO_2$ solid phase.

Another important goal of this series of articles is to clarify the general behavior of RIPI model and its dependence on SPP, system size and boundary conditions (periodic vs. isolated, i.e. finite crystals with surface surrounded by vacuum).

## 2. Methodology

Since we are interested in MD-simulation of the thermophysical properties of fuel, rather than static calculations, we have investigated only pair potentials of rigid ions without the shells, as the shell-core model requires 4 times more computational time and a much smaller time step for integration of the shells dynamics.

In addition to the widely used potentials Walker-81 [5], Busker-02 [6], Morelon-03 [8] and to the potentials Yamada-00 [9], Basak-03 [10], Arima-05 [11], which were assessed in the review of Govers et al. with inaccurate parameters, we compared four new SPPs for $UO_2$ and three SPPs for $PuO_2$. All these SPPs can be divided into 4 groups by the methods of parameterization and experimental data used as reference:

- Potentials Walker-81 [5], Busker-02 [6], Nekrasov-08 [18], Goel-08 [7] were obtained in the harmonic oscillators approximation from the elastic properties at zero temperature (note that Busker and Goel had parameterized the potentials for the shell-core model but suggested using the same parameters for MD with the rigid ions approximation).
- Potentials Morelon-03 [8] were obtained using the lattice statics from the formation energies of Frenkel and Schottky point defects.
- Potentials Yamada-00 [9], Basak-03 [10], Arima-05 [11] were obtained using MD-simulation from the low-temperature (T < 2000K) evolution of bulk modulus and lattice constant.
- Finally, our set of potentials MOX-07 [12] [19] and Yakub-09 [20] [21] (we refer to the latter works because in the original article [13] the coefficients of SPP were given incorrectly) were fitted to the evolution of lattice constant in the whole range of temperatures (from 300K up to melting point) by MD-simulation.

We fitted MOX-07 to the experimental thermal expansion behavior, as it is known with good accuracy over a wide temperature range (up to 2900K for $UO_2$, and up to 1800K for $PuO_2$ [22]). Our preliminary theoretical analysis of this dependence in the self-consistent field approximation showed that it is explicitly determined by the values of potential derivatives of at least until the fourth order. Although derived analytical formulae did not have a quantitative accuracy for the calculations at high temperatures, however, they qualitatively predicted an essential sensitivity of thermal expansion to the coefficients of SPP. As a result, reproducing thermal expansion by MD-parameterization with an accuracy of 0.005Å (less than 0.1%) in the range 900K < T < 2800K for $UO_2$ and in the whole temperature range for $PuO_2$ provided a quantitative reproduction of a wide spectrum of other experimental data (see the results below and the original papers [12] [19]).

The coefficients of all the potentials are given in Tables 1, 2 and 3. Unlike other SPPs, the short range "anion-anion" interaction of Morelon-03 is set piecewise, while Goel-08 has it with zero dispersion term. Short range "cation-anion" interaction of Yamada-00, Basak-03 and Yakub-09 includes covalent Morse term. Finally, the potentials Yamada-00, Basak-03, Arima-05, Yakub-09 and Goel-08 have "cation-cation" exponential repulsion term.

Note that only three of the considered SPPs (Yamada-00, Arima-05 and MOX-07) include the consistent "anion-cation" and "cation-cation" potentials for $PuO_2$ allowing simulation of MOX-fuel $(U,Pu)O_2$. We devoted a separate article to their comparison (including most recent SPP of Tiwary et al [17]).

We carried out all MD-simulations on graphics processors (GPU) using NVIDIA CUDA technology, which gave us speedup of 100–1000 times (see details in [23] [24]). In order to avoid surface effects, we used the periodic boundary conditions (PBC) and Ewald summation of long-range ionic interactions with minimum image convention cutoff in real space $R_{cutoff} = X/2$ [25], Ewald parameter $W = 2\pi/X$ and high-accuracy cutoff in reciprocal space $K_{cutoff} = 6(2\pi/X)$ [26], where X is instantaneous (i.e. variable with barostat fluctuations) supercell edge length.

In order to integrate Newton's equations of motion, we used second-order Verlet method (with time step of 5 femtoseconds), because it is high-accuracy and time-reversible (i.e. symplectic), which ensures the conservation of energy, impulse and momentum. The simulations were carried out under isochoric (NVT) and isobaric (NPT) conditions (at constant number of ions, temperature and volume or pressure), controlled by the quasi-canonical dissipative thermostat and barostat of Berendsen with a relaxation time of 1 picosecond. The equilibration time was 10 ps, and measured values were averaged over the next 20 ps (total of 6000 MD steps).

Lattice constant (L) and enthalpy (H) of the system were measured in NPT-simulations at zero pressure (neglecting standard atmospheric pressure of ~0.1 MPa). Isobaric heat capacity ($C_P$) and linear thermal expansion coefficient (LTEC) were calculated by their numerical differentiation, because the values calculated from the fluctuations depend on MD-algorithms of integration, barostat and thermostat.

In order to get the bulk modulus (B), we measured pressure in NVT-simulations for several specified values of the volume and approximated these measurements linearly as $P(T) = x + y/V(T)$ for each temperature. Differentiating this expression and putting the result $dP/dV = -y/V^2$ into the bulk modulus definition $B = -VdP/dV$, we found its values by the formula $B = y/V$.

We calculated isochoric heat capacity ($C_V$) using $C_P$ and lattice dilation term, given by the well-known thermodynamic relation $C_D = C_P - C_V = 9\alpha^2 BTV_M$ (where $\alpha$ is the LTEC, B is the bulk modulus, T is the temperature, $V_M = N_A L^3/4$ is the volume of one mole of $UO_2$ molecules).

We examined the dependence of model thermophysical characteristics on the system size in the range from 96 (2x2x2 FCC cells) to 12000 (10x10x10 FCC cells) ions. In this case, with half of the SPPs having short-range term "cation-cation" (Yamada-00, Basak-03, Arima-05, Goel-08 and Yakub-09), λ-peak of LTEC and $C_P$ dependences is moved to lower temperatures with increasing system size. However, as seen in Fig. 1, even for these potentials the system of 1500 ions is sufficient, while for the rest SPPs the most noticeable differences are seen only between systems of 96 and 324 ions. Therefore, we give results for the system of 1500 ions in the charts and comparison of values for several sizes in Table 4.

In order to achieve high precision results, for each SPP the temperature dependences were measured with a step of 1K (for systems of 324, 768 and 1500 ions) up to the corresponding melting point, then validated on larger systems (2592, 6144 and 12000 ions) with coarser temperature steps (2K, 5K, 10K). Using such a small steps in temperature has several advantages. First, it allows one to control the random errors, especially in areas with a rapid change in values (e.g. close to phase transitions). Secondly, it allows detecting discontinuities in the curves, even when they are accompanied by a small jump in the measured quantities. In particular, it allows one to detect reliably presence or absence of first/second order phase transitions. Thirdly, it allows obtaining the exact shape of the temperature dependences and their derivatives, in particular, the λ-peaks. It is obvious that one cannot get a narrow λ-peak of width 20–50K (expected in some phenomenological hypotheses [13] [22]) when using a step of 100K.

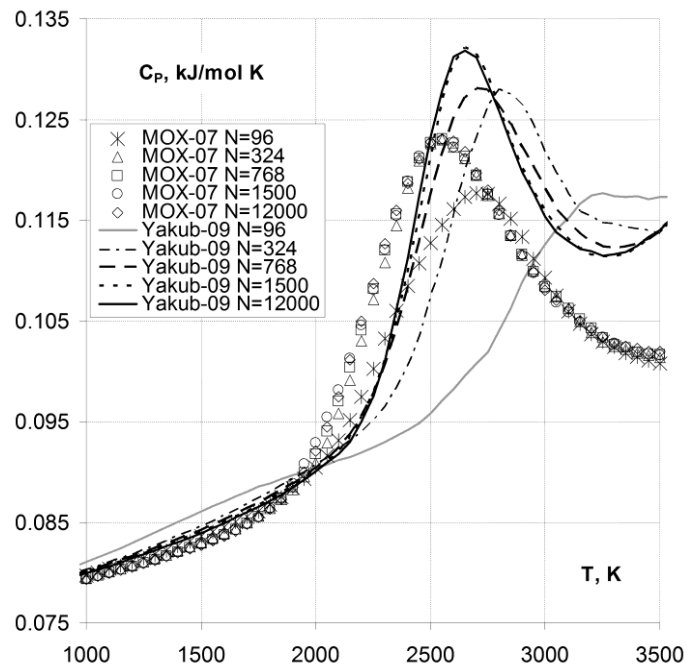

FIG. 1. Influence of system size on isobaric heat capacity.

However, in order to smooth the curves in the charts we averaged lattice constant, enthalpy and bulk modulus over the interval of ±25K; LTEC and heat capacity were averaged over the interval of ±100K twice. Simulations where melting began, were excluded from the averaging, because otherwise they would raise the measured lattice constant and enthalpy. For all SPPs (except Yamada-00) we obtained the following estimates of the uncertainty (maximum absolute error of arithmetic mean): $0.3*10^6$/K in the LTEC charts, 0.0007 kJ/(mol*K) in the heat capacity charts (the largest errors were obtained in the regions of λ-peak and melting).

The curves for Yamada-00 potentials were measured with integration time step of 1 fs (instead of 5 fs) and plotted in figures without averaging (except LTEC and heat capacity) to emphasize their anionic sublattice instability, i.e. multiple instantaneous first-order phase transitions between crystalline and disordered states (see Fig. 2 and discussion in the next section).

In order to measure the melting temperature ($T_{melt}$) for each SPP we conducted a series of simulations at temperatures where melting occurs with a step of 1K and simulation time of 500 ps ($10^5$ MD steps). After that, from the whole series of length 400K we chose a temperature interval of length 10K, which would include at least three melting events (detected by sharp changes in density and enthalpy). Determining the melting point by one such event would be less reliable due to the stochastic kinetic initiation of this phase transition.

Bredig superionic phase transition temperature (premelting of anionic sublattice) was measured from the λ-peak of the isochoric heat capacity in order to exclude the lattice dilation term influence and rounded to the nearest multiple of 10K.

Experimental data in Fig. 3–9 and Table 4 are marked with the prefix "exp" and data recommended in reviews with the prefix "rec".

### 2.2. Anomalies of Yamada-00 potentials

As one can see from Fig. 2 anionic sublattice instability of Yamada-00 is more pronounced for smaller systems (e.g. 1400–3400K for system of 324 ions), but first-order phase transition persists even in a system of 12000 ions. Fig. 3, 4 and 7 shows that lattice constant and enthalpy for this SPP suffer from discontinuous oscillations in the range 2000–2600K, i.e. first-order phase transitions happens spontaneously due to the aforementioned anionic sublattice instability. Obviously, the numerical derivatives (LTEC and heat capacity) may take arbitrarily large values depending on the step in temperature (see curves with 0.1K and 1K step in Fig. 8), so their quantitative discussion in this interval does not make sense. Conversely, in MD-simulations with the other reviewed potentials both lattice constant and enthalpy increase continuously with temperature in the region of superionic transition, and their derivatives have a smooth λ-peak. However, in simulations under a large negative pressure of –20 GPa similar instability was observed with Walker-81, Busker-02, Goel-08, while the other SPPs remained stable.

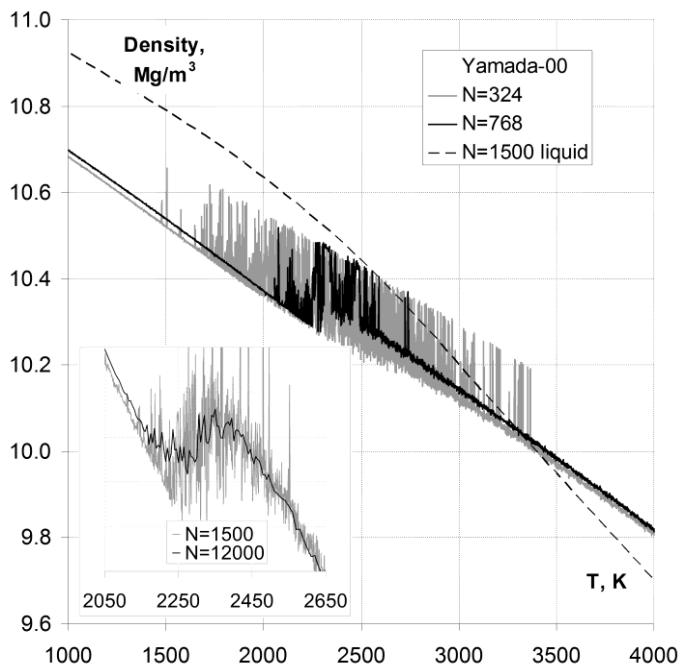

FIG. 2. Anomalies of Yamada-00: lattice instability on small systems, first-order phase transition (big systems close-up in subfigure), inverse density ratio of 3 phases (liquid > superionic > crystal).

Besides, Yamada-00 SPP is oscillatory not only in NPT PBC simulations (i.e. Ewald summation of pair interactions with minimum image convention cutoff varied by barostat), but also in NVT PBC simulations (with fixed cutoff) as seen from bulk modulus curve in Fig. 6. Moreover, during NPT simulations under isolated (non-periodic) boundary conditions (i.e. direct summation of pair interactions without any cutoff) with Yamada-00 SPP the anionic sublattice is disordered at all temperatures, unlike with the rest of SPPs. Hence this instability is most likely inherent property of Yamada-00 potentials.

In the simulations of other authors with Yamada-00 SPP the discontinuities and first-order phase transition with infinitely-high peaks were presumably hidden due to a coarse step in temperature. Peaks of LTEC and $C_P$ in our Fig. 5 and 8 are also finite but due to averaging of discontinuities. LTEC of Yamada-00 has a "λ-pit" instead of λ-peak near 2300K, because as seen in Fig. 2 superionic phase (with disordered anionic sublattice) is denser than crystal solid phase (moreover, the supercooled liquid at that temperature has even greater density). It is interesting that such density ratio of these three phases is not unique: we have been able to reproduce it with MOX-07 potentials under pressure of 20 GPa, although only a first-order phase transition with "λ-pit" of LTEC but without anionic sublattice instability was observed in this case.

### 3. Results and discussion

### 3.1. Ionicity

First of all, let us compare the coefficient of ionicity (Q) of pair potentials with the recommended estimates. This is the only parameter that determines interaction of ions at distances of more than 2–3 lattice periods, therefore its influence on defect formation and phase transitions can hardly be compensated by other SPP parameters.

Pauling's empirical formula [27] states that $Q = 1 - \exp(-(\Delta x)^2/4)$, where $\Delta x$ is difference of the electronegativities of metal and oxygen, the recommended values of which are as follows: 3.44 for oxygen, 1.38 for uranium and 1.28 for plutonium [28]. Thus $\Delta x = 3.44 - 1.38 = 2.06$ and $Q = 0.654$ for $UO_2$; $\Delta x = 3.44 - 1.28 = 2.16$ and $Q = 0.689$ for $PuO_2$.

These estimated values of Q are within 5% from the value of 0.68623 for our MOX-07 potentials, which are fitted for MD simulation of MOX-fuel $(U,Pu)O_2$. Note, however, that electronegativity of elements is not measured experimentally, but calculated in different ways using binding energy and other thermodynamic data [29], so one could not expect the exact coincidence.

Walker-81 and Busker-02 potentials have formal ion charges +4 and –2 (*a priori* value Q = 1). The coefficients of ionicity in the potentials Yamada-00, Basak-03, Arima-05, and Goel-08 were also set *a priori*, but different from the unit. Finally, Nekrasov-08, Morelon-03 and Yakub-09 while have the parameter Q fitted, nevertheless, obtained values of 0.95425, 0.806813 and 0.5552 are far from the above estimates (however in work [13] authors noted closeness of their fitted parameter Q to the value of 0.555 given by Pauling's formula with $\Delta x = 3.5 - 1.7 = 1.8$ instead of 2.06).

### 3.2. Thermal expansion and bulk modulus

The temperature dependences of simulated thermophysical parameters are shown in Fig. 3–9 up to 5700K (to the region of Busker-02 λ-peaks). The chart of lattice constant deviation is shown up to 3100K, and the chart of enthalpy is shown in the range 1300–3100K, in order to emphasize differences with recommendations from the most recent IAEA review of experimental data on reactor materials [22].

The curves for each SPP are plotted up to the corresponding melting points (see Table 4). The cause of too high melting temperatures (compared with experimental values [22] [30]) is that crystals which are MD-simulated under PBC melt in a superheated state (spinodal condition) due to the lack of surface (or other defects in the cationic sublattice). In order to overcome this effect, some authors [15] [21] have measured the temperature of equilibrium of two-phase crystal-melt systems under PBC (binodal condition). But we believe that melting simulation of nanoscopic crystals with surface (which are finite and surrounded by vacuum, i.e. under isolated boundary conditions) would be more correct, and have devoted a separate article to this issue [31]. But here we note that all the model values of $T_{melt}$ exceed the recommended value of 3140±20K (in inert atmosphere) by more than 20%, and only three SPPs (Goel-08, Yakub-09 and MOX-07) have $T_{melt}$ ~4000K or less. Also Table 4 shows that the melting point is weakly dependent on system size: for most SPPs obtained values differ by less than 50K, starting with a system of 324 ions.

Fig. 3 and 4 shows thermal expansion L(T) and its deviation ΔL(T) from IAEA recommendations [22], where one can observe the following features:

- with Walker-81 L(T) always lies below the experiment by 0.13–0.24Å, and with Goel-08 – always above (up to ~0.038Å at 1900K), approaching ~0.01Å only at outermost points (at 300K and 3050K);
- with Busker-02, Nekrasov-08, Yamada-00 the deviation does not exceed 0.01Å in the ranges from 300K to 1150K, 1700K, 2000K respectively, and at 3150K it reaches the maximum values of 0.14Å, 0.1Å and 0.11Å;
- with Arima-05, Basak-03, Morelon-03 the deviation does not exceed 0.01Å in the ranges from 300K to 2300K, 2650K, 2550K respectively, and at 3150K it reaches the maximum values of 0.047Å, 0.043Å, 0.041Å (twice as small as previous group);
- finally, with Yakub-09 the deviation does not exceed 0.01Å in the range 300–2450K, but unlike most of the potentials its deviation has a maximum value of 0.014Å already at 2700K; with MOX-07 the deviation does not exceed 0.01Å in the widest range – up to 2900K, but at 3150K it reaches the value of 0.028Å (1.5 times less than Morelon-03, but 2 times more than Yakub-09).

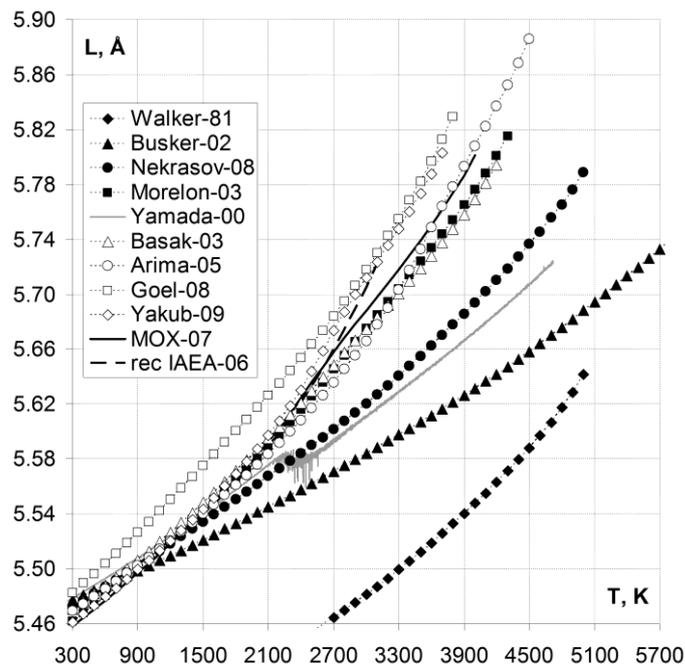

FIG. 3. Temperature dependence of lattice constant.

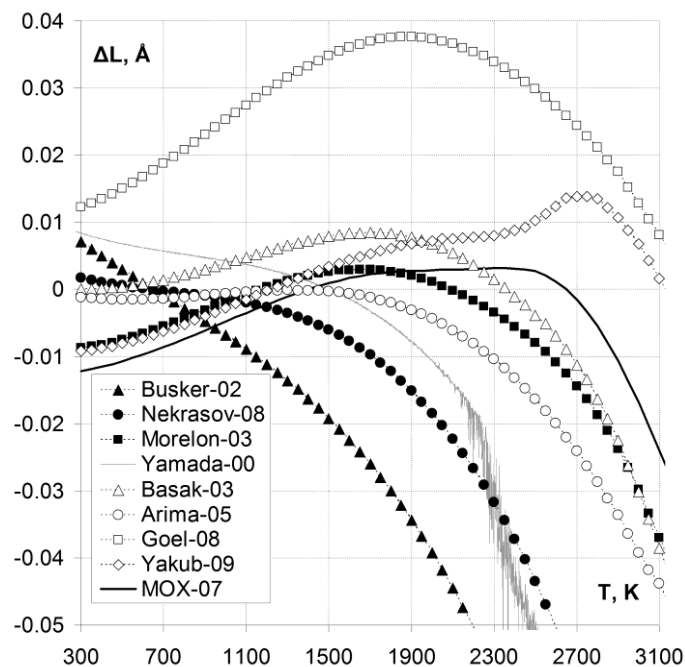

FIG. 4. Deviation of lattice constant from IAEA recommendation.

Fig. 5 shows that Walker-81, Yamada-00, Busker-02, Nekrasov-08 and Goel-08 have a nondecreasing temperature dependence of LTEC; Basak-03 and Morelon-03 demonstrate a weak λ-peak and only Arima-05, Yakub-09, MOX-07 provide an obvious λ-peak. Note that MOX-07 reproduces the experimental curve in the range 1700–2500K, i.e. up to the superionic transition region, divergence after which is due to the lower lattice constant at T > 2700K. In addition, it can be seen that above 3100K its LTEC rises again (tending to infinity at the melting point of ~4000K), and if one mentally moves this section of the curve by ~850K to the left (as if melting occurred at 3150K), then the model and recommended dependences coincide. Similar behavior is shown by Yakub-09: its LTEC concurs the experimental curve in the range 1900–2750K, then sharply decreases just after the superionic transition (in spite of higher lattice constant at T > 2700K), but raises above 3200K tending to infinity at the melting point near 3800K. Finally, Arima-05 goes near recommendation up to 1500K, but has substantially higher temperatures of the λ-peak and melting: 3530K and 4550K correspondingly.

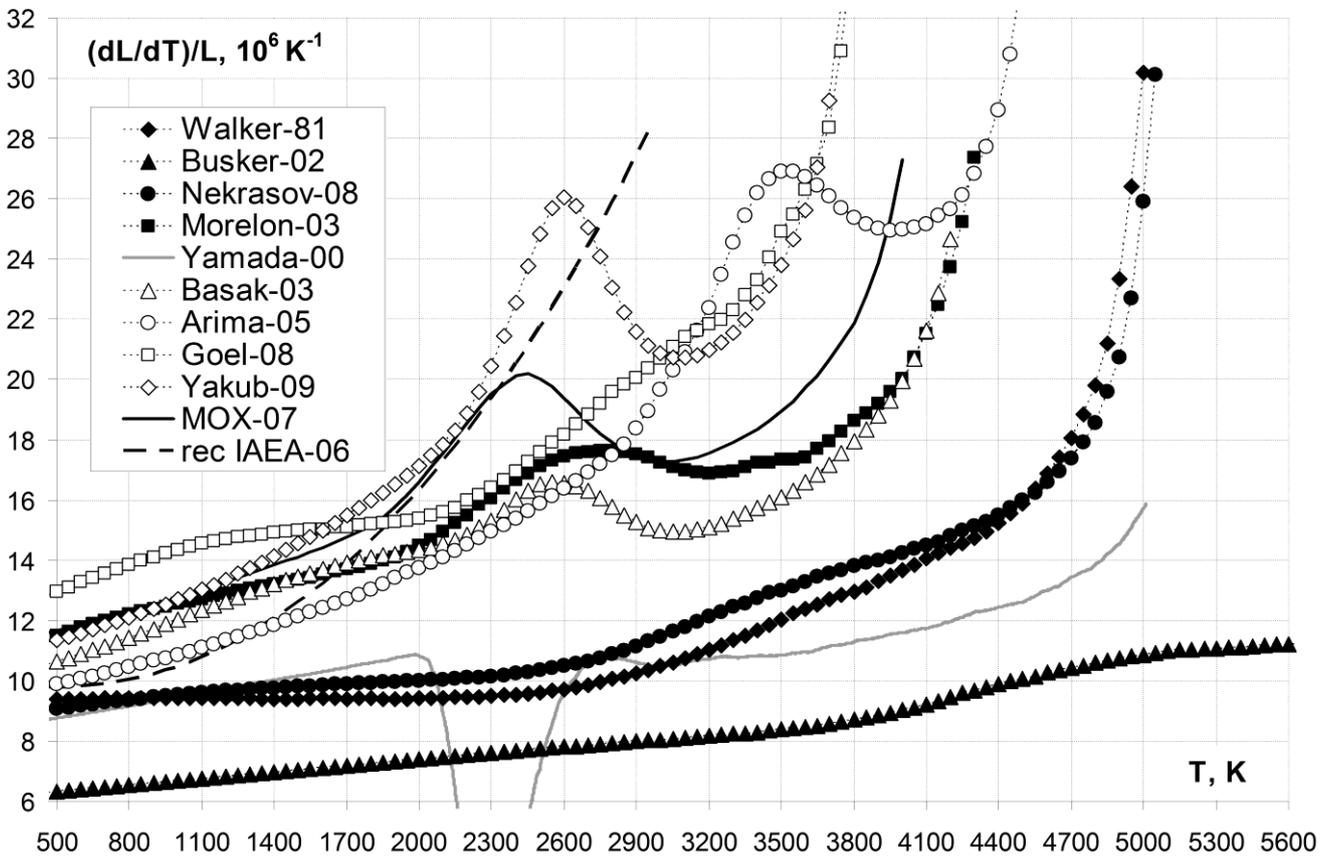

FIG. 5. Temperature dependence of linear thermal expansion coefficient.

Unfortunately, the review [22] does not consider bulk modulus (or isothermal compressibility, which is inverse to it), so we should base comparison on the experimental data of Hall [32] and Browning [33] (which are available in a narrow interval T<1600K, and their values differ by 1.5 times) and on the older recommendation of Martin [34]. Most of the model curves in Fig. 6 are S-shaped (though a curvature is almost unnoticeable with Busker-02, Morelon-03 and MOX-07). However, Arima-05, MOX-07 and Yakub-09 have additional hump (less noticeable with MOX-07) near their superionic transition temperatures.

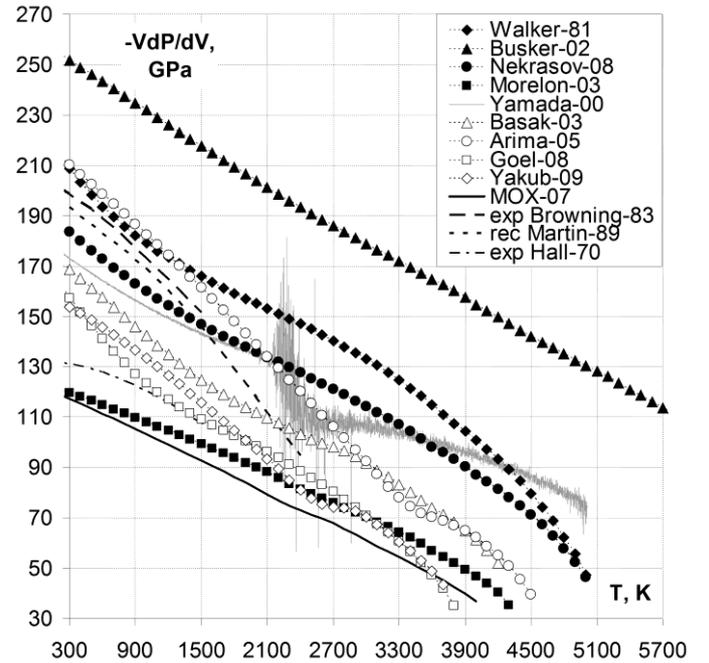

FIG. 6. Temperature dependence of bulk modulus.

Since experimental data of Browning were used as the reference in fitting of most SPPs, only the results for Morelon-03 and MOX-07 can be considered as independent estimates. Fig. 6 shows that these potentials reproduce slope and position of Hall's experimental data. Besides, extrapolations of Hall's and Martin's data intersect with each other and with the model curve of MOX-07 on the value of 60 GPa at the experimental melting temperature ~3150K, while Morelon-03 has a smaller slope and intersect Goel-08 and Yakub-09 on the value of 70 GPa at 3150K. Notice that most of the model curves at their melting points reach values in a narrow range 40±5 GPa, with the exception of Yamada-00's value of 76 GPa at 5000K and Busker-02's value of 65 GPa at 7110K.

### 3.3. Enthalpy and specific heat capacity

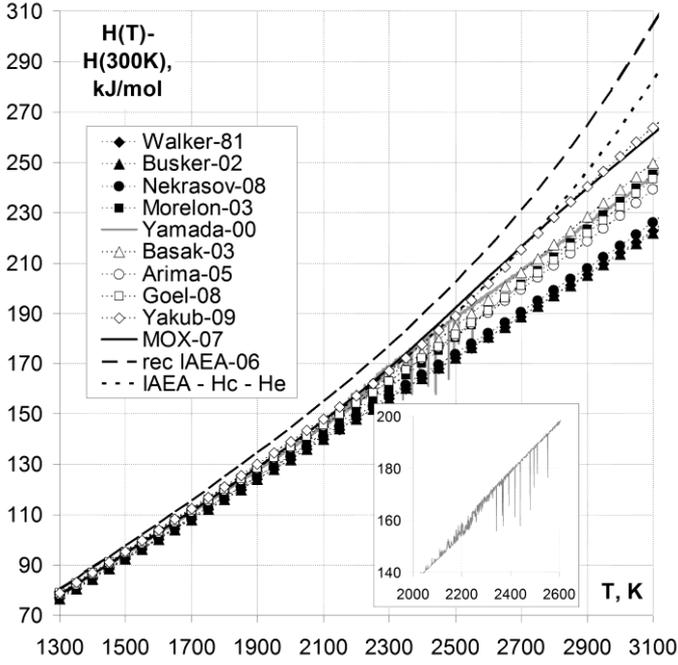

FIG. 7. Temperature dependence of enthalpy (subfigure shows close-up of Yamada-00 anomalies).

Evolution of enthalpy for all SPPs coincide with the IAEA recommendation at temperatures less than 1300K, so this range is excluded from Fig. 7 in order to emphasize the differences. Since MD-simulation under PBC in RIPI approximation does not allow the formation of either electronic defects (in particular, high-temperature polaronic disorder of type: $2U^{4+} \leftrightarrow U^{3+} + U^{5+}$) or Schottky defects (i.e. trivacancies with molecules rising to the surface), then in order to estimate their contributions to enthalpy we used empirical equations from the work [9]: $H_e = 256*\exp(-10790/T)$ kJ/mol and $H_c = 0.00000146*T^2$ kJ/mol. Fig. 7 shows that the model curves do not agree with the recommended dependence at high temperatures even with these contributions subtracted (see the curve "IAEA–Hc–He"). However, the results for Yakub-09 and MOX-07 behave much better, reproducing the recommended curve up to 2800K with the maximum deviation at 3150K being only 8% (unlike 2200K and 12% with other SPPs).

Fig. 8 shows isobaric heat capacity. The chart was cut below 500K, as MD-simulation in RIPI approximation does not consider quantum mechanical effects and cannot reproduce the sharp decrease of the experimental heat capacity at lower temperatures. However, this behavior can be obtained from phonon spectrum calculations. One can see dependence on temperature step for Yamada-00 (curves with 0.1K and 1K step) due to non-differentiability of the enthalpy discontinuities in the range 2000–2600K. It is seen that Yakub-09 and MOX-07 reproduce the recommended curve "IAEA–Cc–Ce" (where $C_e$ and $C_c$ are derivatives of the empirical contributions $H_e$ and $H_c$) until the superionic transition region near 2600K (while the rest of SPPs coincide only up to 1800K).

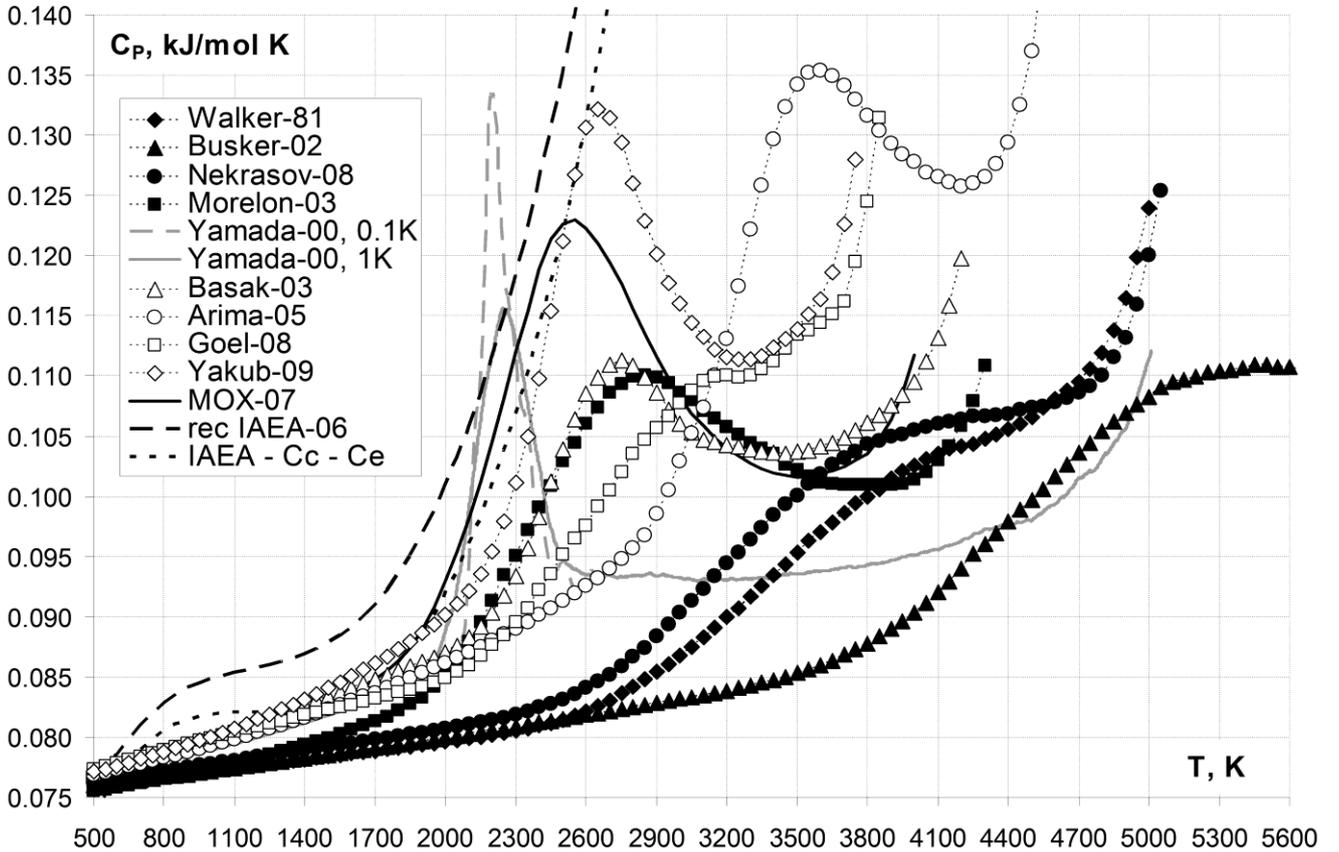

FIG. 8. Temperature dependence of isobaric heat capacity.

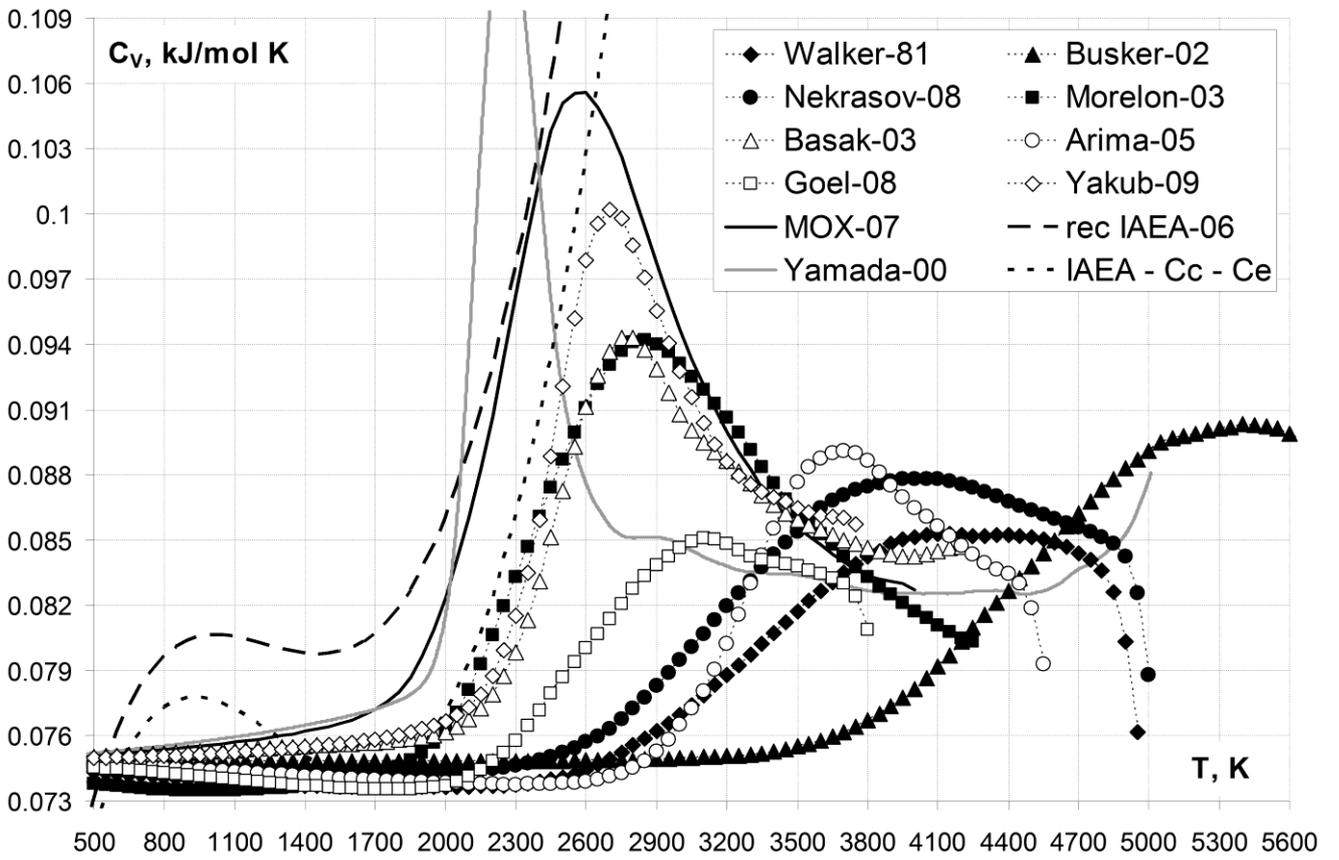

FIG. 9. Temperature dependence of isochoric heat capacity.

Fig. 9 shows the comparison of isochoric heat capacity curves of various SPPs with the reference curve named IAEA-06, which is derived from the IAEA recommendations [22] of L, LTEC, $C_P$ and the Martin's recommendation [34] of bulk modulus. Note that in this chart λ-peaks are clearer than in isobaric heat capacity charts, especially for Walker-81, Busker-02 and Nekrasov-08 SPPs.

Note also that all the model curves in Fig. 9 decrease after superionic transition until the corresponding melting point. Therefore, the growth of isobaric heat capacity after superionic transition seen in Fig. 8 is determined exclusively by the lattice dilation term $C_D$. In particular, Arima-05 SPP has the maximum λ-peak value in Fig. 8, because of the high LTEC peak in Fig. 5, but in Fig. 9 its λ-peak is 2.5 times lower than that of MOX-07.

Table 4 shows that $C_V$ λ-peak temperature changes with system size by 100–300K (except almost independent Morelon-03 and MOX-07 potentials) but saturates already at 1500 ions. Only three of the SPPs (Basak-03, Yakub-09, MOX-07) have a temperature of λ-peak close to the current IAEA recommendation 2670K [22] and the Ralph's experimental value of 2610K [35]. Whereas with 5 of the 10 SPPs it is above the experimental melting point and with Yamada-00 at T ~2250K the first-order phase transition is observed with inverse density jump (superionic phase is denser than the crystal).

We associate manifestation of λ-peak in LTEC, $C_P$ and $C_V$ charts with the saturation of anti-Frenkel defects concentration (since each defect increases the enthalpy and lattice constant). The known estimates of band gap in $UO_2$ do not exceed 2 eV [3], which is significantly less than anti-Frenkel defect formation energy ~4 eV [3]. Therefore, polarons (which are absent in our model, but partially compensated by $C_e$) are beginning to affect the experimental dependences at lower temperatures and manifest itself as the higher slope of IAEA recommendation compared with the model curves at 1000–1800K. On the other hand, Schottky defects, which also affect the experimental dependences, have higher formation energy ~7 eV [3], and therefore they should occur at higher temperatures. Their absence in our model leads to a difference between the model curves and IAEA recommendation at temperatures above 2700K, which is partially compensated by $C_c$. At temperatures close to melting the cationic Frenkel defects, presumably, begin to form (instead of Schottky defects which require some kind of surface), contributing to the model dependences. However, as seen in Fig. 9, due to their highest formation energy 15–23 eV [3] [14] this contribution is not large enough to provide the growth of curves, and only slows their fall.

The discrepancy of the model and experimental curves after the superionic transition observed in Fig. 8 and 9 may have several reasons: inaccuracy of the empirical contributions $C_e$ and $C_c$, difference in the processes of cationic sublattice disordering or divergence of the original temperature dependences (Fig. 3–7) in particular high melting temperatures of MD simulations under PBC. In order to clarify the situation, we present the corresponding analysis of various contributions to simulated heat capacity in our other article [36].

TABLE 1. Ionicity (Q) and short-range parameters of the potentials in Buckingham form
$X*\exp(-Y*R) - Z/R^6$.

| SPP | Q | X-- eV | Y-- 1/Å | Z-- eV*Å$^6$ | X+- eV | Y+- 1/Å | X++ eV | Y++ 1/Å |
|---|---|---|---|---|---|---|---|---|
| UO$_2$ Walker-81 | 1 | 50259.3 | 6.54236 | 72.6534 | 873.327 | 2.477 | – | – |
| UO$_2$ Busker-02 | 1 | 9547.96 | 4.562 | 32 | 1761.78 | 2.806 | – | – |
| UO$_2$ Morelon-03 | 0.806813 | * | * | * | 566.498 | 2.37778 | – | – |
| UO$_2$ Nekrasov-08 | 0.95425 | 50259.3 | 6.54236 | 72.6534 | 873.327 | 2.477 | – | – |
| UO$_2$ Goel-08 | 0.725 | 1822 | 3.53257 | – | 1822 | 3.21143 | 1822 | 2.94381 |
| UO$_2$ Arima-05 | 0.675 | 978.718 | 3.01205 | 17.3544 | 55892.6 | 4.95050 | 2.48128e+13 | 13.8889 |
| PuO$_2$ Arima-05 | 0.675 | 978.718 | 3.01205 | 17.3544 | 57425.2 | 5.03778 | 2.80460e+14 | 15.3846 |
| **UO$_2$ MOX-07** | **0.68623** | **50211.7** | **5.52** | **74.7961** | **873.107** | **2.78386** | – | – |
| **PuO$_2$ MOX-07** | **0.68623** | **50211.7** | **5.52** | **74.7961** | **871.790** | **2.80788** | – | – |

\* – see in Table 2.

TABLE 2. "Anion-anion" short-range parameters of SPP Morelon-03.

| Distance, Å | Short range term, eV |
|---|---|
| 0 < R < 1.2 | $11272.6 * \exp(-7.33676 * R)$ |
| 1.2 < R < 2.1 | $-27.2447 * R^5 + 246.435 * R^4 - 881.969 * R^3 + 1562.22 * R^2 - 1372.53 * R + 479.955$ |
| 2.1 < R < 2.6 | $-3.13140 * R^3 + 23.0774 * R^2 - 55.4965 * R + 42.8917$ |
| R > 2.6 | $-134 / R^6$ |

TABLE 3. Ionicity (Q) and short-range parameters of the potentials in "Buckingham with Morse" form
$X*\exp(-Y*R) - Z/R^6 + G*((\exp(-H*(D - R)) - 1)^2 - 1)$.

| SPP | Q | X-- eV | Y 1/Å | Z-- eV*Å$^6$ | X+- eV | G+- eV | H+- 1/Å | D+- Å | X++ eV |
|---|---|---|---|---|---|---|---|---|---|
| UO$_2$ Yamada-00 | 0.6 | 2345.90 | 3.125 | 4.14572 | 1018.46 | 0.780930 | 1.25 | 2.369 | 442.161 |
| PuO$_2$ Yamada-00 | 0.6 | 2345.90 | * | 4.14572 | 5329.83 | 0.564005 | 1.56 | 2.339 | 32606.8 |
| UO$_2$ Basak-03 | 0.6 | 1633.01 | 3.0579 | 3.94880 | 693.651 | 0.577190 | 1.65 | 2.369 | 294.641 |
| UO$_2$ Yakub-09 | 0.5552 | 883.12 | 2.9223 | 3.996 | 432.18 | 0.5055 | 1.864 | 2.378 | 187.03 |

\* – exponents for the plutonium dioxide SPP are different: Y-- = 3.125, Y+- = 4.16667, Y++ = 6.25.

TABLE 4. Size dependence of phase transitions temperatures.

| SPP | Melting temperature, K | | | | C$_V$ λ-peak temperature, K | | | |
|---|---|---|---|---|---|---|---|---|
| | N=324 | N=768 | N=1500 | N=12000* | N=324 | N=768 | N=1500 | N=12000* |
| Walker-81 | 4900 | 4990 | 4980 | 5000 | 4300 | 4070 | 4160 | 4080 |
| Busker-02 | 6950 | 7110 | 7100 | 7100 | 5460 | 5350 | 5410 | 5340 |
| Nekrasov-08 | 4950 | 5050 | 5030 | 5040 | 4140 | 3940 | 4000 | 4000 |
| Morelon-03 | 4270 | 4260 | 4270 | 4260 | 2890 | 2900 | 2840 | 2860 |
| Yamada-00 | 4960 | 5000 | 5010 | 5000 | ** | ** | 2240 | 2230 |
| Basak-03 | 4170 | 4200 | 4200 | 4200 | 3060 | 2910 | 2770 | 2740 |
| Arima-05 | 4520 | 4550 | 4550 | 4550 | 3820 | 3730 | 3710 | 3680 |
| Goel-08 | 3840 | 3830 | 3840 | 3840 | 3370 | 3240 | 3110 | 3140 |
| Yakub-09 | 3720 | 3760 | 3750 | 3750 | 2860 | 2740 | 2700 | 2720 |
| MOX-07 | 4000 | 3990 | 4010 | 4000 | 2580 | 2580 | 2570 | 2590 |
| rec IAEA-06 | [22] 3140±20*** | | | | [22] 2670 | | | |
| Experiments | [30] 3150±20*** | | | | [35] 2610 | | | |

\* – by measurements with temperature step of 10K instead of 1K;

\*\* – non-differentiable anionic sublattice instability; \*\*\* – in inert atmosphere without oxygen.

## 4. Conclusions

Compared with previous works on MD simulation of uranium dioxide ($UO_2$), the use of graphics processors (GPU) and NVIDIA CUDA technology has allowed performing a large amount of numerical experiments for 10 sets of pair potentials (SPP) in a wide range of temperatures (from 300K up to melting point) with a step of 1K which guaranteed high accuracy of the temperature dependences charts for characteristic thermophysical quantities.

Due to the high-precision measurements we revealed the λ-peak of heat capacity with each of 10 considered SPPs. Although λ-peaks were not always visible or unclear in a $C_P$ chart, they are unambiguously characterized in a $C_V$ chart.

The best reproduction of considered $UO_2$ properties is demonstrated by two recent SPPs MOX-07 [12] and Yakub-09 [20], which both had been fitted to the recommended thermal expansion in the range of temperatures 300–3100K. They agree with the experimental data better at temperatures above 2500K than the widely used SPPs Basak-03 [10] and Morelon-03 [8], which were chosen as the best in the review of Govers et al. [14] [15] (because of MOX-07 and Yakub-09 later publication). The divergence of model and recommended dependences above 2700K is presumably due to absence of Schottky defects formation in MD simulations without surfaces.

Less adequate behavior is shown by the Arima-05 [11] and Goel-08 [7] potentials, but the worst were the oldest SPPs: Busker-02 [6], Yamada-00 [9] and Walker-81 [5] (including its "ionicity" modification Nekrasov-08 [18], which corrects the lattice constant).

While investigating Yamada-00 potentials we revealed an interesting anomalies: anionic sublattice instability and corresponding first-order phase transition with inverse density jump (superionic phase is denser than the crystal), not found by other authors [9] [13] [15] probably due to coarse temperature step of their simulations. Instead of a continuous anionic disordering Yamada-00 have a region of metastable coexistence of two phases with spontaneous step-wise changes of characteristics (lattice constant, enthalpy and, as will be shown in the next article, anion self-diffusion coefficient).

When the article was ready, in the process of review, we were asked to assess the new shell-core potentials of Read and Jackson [37] (Read-10) and *ab initio* potentials of Tiwary et al. [17]. We examined Read-10 SPP in the approximation of rigid ions, and its results with almost linear temperature dependences, melting point of ~6600K and $C_V$ λ-peak temperature of ~4700K are placed between the results of Busker-02 and Nekrasov-08. Therefore (from results of Busker-02 and Read-10 SPPs) one can see that shell-core potentials with formal charges are unsuitable for use without shells. In contrast, for example, to the shell-core potentials Goel-08 with ionicity of 0.725, which provide more adequate behavior in approximation of rigid ions. Regarding *ab initio* potentials of Tiwary et al., they are going to be considered in our future article on MD-simulation of PuO2 and MOX (including the corresponding results for $UO_2$).

The following articles in this series will embrace examination of melting [31], superionic transition [36] and diffusion in both quasi-infinite periodic crystals and finite nanoscopic crystals with surface (surrounded by vacuum), as well as simulation of plutonium dioxide and MOX fuel of $(U,Pu)O_2$ type.